\documentclass[seceq]{ptptex}
\usepackage{bm}
\usepackage{graphicx,color}
\def\G{g}
\begin{document}



\pubinfo{Vol.~113, No.~2, February 2005}
\def\ptype{p}                        
\def\ptpsubject{3}                    
\notypesetlogo                       

\markboth{W. Hikida et al. }{Self-Force Regularization. II}

\title{A New Analytical Method for Self-Force Regularization. II}

\subtitle{Testing the Efficiency for Circular Orbits}

\author{Wataru \textsc{Hikida},$^{1,2,}$%
\footnote{E-mail: hikida@yukawa.kyoto-u.ac.jp}
Sanjay \textsc{Jhingan},$^{3,}$%
\footnote{E-mail: wtpsaxxj@lg.ehu.es}
Hiroyuki \textsc{Nakano},$^{4,}$%
\footnote{E-mail: denden@sci.osaka-cu.ac.jp}\\
Norichika \textsc{Sago},$^{5,}$\footnote{E-mail:
sago@vega.ess.sci.osaka-u.ac.jp} Misao \textsc{Sasaki}$^{1,}$%
\footnote{E-mail: misao@yukawa.kyoto-u.ac.jp} and Takahiro
\textsc{Tanaka}$^{2,}$\footnote{E-mail: tama@scphys.kyoto-u.ac.jp} }

\inst{
$^1$Yukawa Institute for
Theoretical Physics, Kyoto University,\\ Kyoto 606-8502, Japan
\\
$^2$Department of Physics, Graduate School of Science, Kyoto
University,\\ Kyoto 606-8502, Japan
\\
$^3$Departamento de F\'{\i}sica Te\'orica, Universidad del
Pa\'{\i}s Vasco,\\ Apdo. 644, 48080, Bilbao, Spain
\\
$^4$Department of Mathematics and Physics, Graduate School of
Science,\\ Osaka City University, Osaka 558-8585, Japan
\\
$^5$Department of Earth and Space Science, Graduate School of
Science,\\ Osaka University, Toyonaka 560-0043, Japan
}


\recdate{November 4, 2004}

\abst{In a previous paper, based on the black hole perturbation approach,
we formulated a new analytical
method for regularizing the self-force acting on a particle of small
mass $\mu$ orbiting a Schwarzschild black hole of mass $M$, where $\mu\ll M$.
In our method, we divide the self-force into the $\tilde S$-part and
$\tilde R$-part. All the singular behavior is contained in the $\tilde
S$-part, and hence the $\tilde R$-part is guaranteed to be regular.  In
this paper, focusing on the case of a scalar-charged particle for
simplicity, we investigate the precision of both the regularized $\tilde
S$-part and the $\tilde R$-part required for the construction of
sufficiently accurate waveforms for almost circular inspiral orbits.
We calculate the regularized $\tilde S$-part
for circular orbits to 18th post-Newtonian (PN) order and investigate
the convergence of the post-Newtonian expansion.
 We also study the convergence of the remaining
$\tilde{R}$-part in the spherical harmonic expansion. We find that
a sufficiently accurate Green function can be obtained by keeping
the terms up to $\ell=13$.}

\maketitle

\section{Introduction}

We are at the dawn of gravitational wave astronomy. Several
ground-based interferometric gravitational wave detectors are in
various stages of development~\cite{LIGO,TAMA,GEO,VIRGO}. R\&D
studies of a space-based gravitational wave observatory project,
the Laser Interferometer Space Antenna (LISA)~\cite{LISA}, which
will observe gravitational waves in the {\rm mHz}-band, are in
rapid progress. There is also a proposal for a DECi hertz
Interferometer Gravitational wave Observatory
(DECIGO/BBO)~\cite{DECIGO,BBO}. This will be a laser interferometer
gravitational wave antenna in space sensitive to frequencies $
f\sim 0.1\ {\rm Hz}$, near which all the currently planned detectors are
insensitive.

Among the promising targets for LISA are binary systems
consisting of a supermassive black hole ($M \sim 10^{5-8}
M_{\odot}$) and a compact object of near solar mass ($\mu \sim 1$--$10
M_{\odot}$). Such gravitational wave sources could provide the first
high-precision tests of general relativity in very strong
gravitational regimes. It is expected that gravitational waves will be
observed from the inspiral stage of these binary systems.
Because the typical observation time of the inspiral
stage is very long, ranging from several months to several
years, high-order relativistic corrections should be observable. The
standard post-Newtonian approximation~\cite{PN-latestreview} seems
inappropriate for this type of binary, i.e. with an extremely
large mass ratio, because the extension of the PN approximation to 
higher post-Newtonian orders is not straightforward in this case.
However, in such a system, there is another natural perturbation
parameter, $\mu/M \sim 10^{-6}$. Therefore, the black hole perturbation
approach is suited for such a binary system~\cite{MSSTT,Sasaki-Tagoshi}.

We consider the self-force acting on a particle of small mass
$\mu$ orbiting a black hole of mass $M$, where $\mu\ll M$.
The central black hole $M$ fixes the metric of the `background' spacetime.
At lowest order in the mass ratio, $(\mu/M)^0$, the particle moves
along a geodesic in the background geometry. Already at this
lowest order, combined with the assumption of adiabatic orbital
evolution, this approach has proven to be very powerful for
evaluating general relativistic corrections to the gravitational
waveforms~\cite{MSSTT,Sasaki-Tagoshi}.
At the next order, however, the spacetime
geometry is perturbed, and the motion of the particle is affected by
its self-force. When we consider the point particle limit, the
full self-force diverges at the position of the particle, and hence
it needs to be regularized. The field generated by the particle 
can be expressed in terms of the retarded Green function.
The retarded Green function can be decomposed into the direct part 
and the tail part, and the properly regularized
self-force is given by the tail part (or the $R$-part) of the
self-field, which is obtained by subtracting the direct part (or the
$S$-part) from the full field. This was first shown by
DeWitt and Brehme~\cite{DB} for the scalar and electro-magnetic cases
decades ago, and rather recently by Mino, Sasaki and Tanaka~\cite{MST},
and Quinn and Wald~\cite{QW} for the gravitational case. An
equivalent but more elegant decomposition of the Green function
was proposed by Detweiler and Whiting~\cite{DW03}, in which the
direct part is replaced by the $S$-part and the tail part by the
$R$-part. The $S$-part is defined so as to vanish when the two
arguments $x$ and $x'$ are timelike,  and to satisfy the same
equation as the retarded Green function. The latter condition
implies that the $R$-part satisfies the source-free homogeneous
equation.

To avoid various technical difficulties associated with the
gravitational case, many previous papers 
focused on the regularization problem of the
scalar radiation reaction force~\cite{Barack:1999zw, Burko:1999zy,
Burko:1999tt, Barack:1999wf, Burko:2000xx, Barack:2000eh,
Barack:2000zq, Burko:2000yx, Nakano:2001kw, Barack:2002mh,
Detweiler:2002gi}. Even in this simplified case of a scalar-charged
particle, however, only special orbits, such as circular or radial orbits,
have been considered, and no systematic method for computing the 
regularized self-force has been given.

Recently~\cite{HJNSST}, we formulated a new analytical method
for regularizing the self-force acting on a particle in the
Schwarzschild spacetime. As is commonly the case
in the self-force regularization problem, our method also 
uses the spherical harmonic decomposition
to obtain a regularized expression for the
self-force~\cite{Barack:1999zw, Burko:1999zy, Burko:1999tt,
Barack:1999wf, Burko:2000xx, Barack:2000eh, Barack:2000zq,
Burko:2000yx, BMNOS, MNS, Barack:2002bt, Barack:2002mh,
Barack:2003mh, Detweiler:2002gi, Diaz-Rivera:2004ik}. The novel point of
our method is that it provides 
a new decomposition of the retarded Green
function in the frequency domain,
which we call the $\tilde S$ and $\tilde R$-parts. This
decomposition guarantees that all the singular behavior is
contained in the $\tilde S$-part. In the black hole perturbation approach,
the Green function is conventionally calculated in the
frequency domain, while the $S$-part to be subtracted is given in
the time domain. We presented a systematic method for
translating the $\tilde S$-part into an expression in the time
domain, and gave an explicit expression for the regularized 
$\tilde S$-part of the self-force, that is, the ($\tilde{S}-S$)-part
for general orbits.

Our method for handling divergences in the $\tilde S$-part relies
on the slow motion approximation, i.e., the post-Newtonian (PN)
expansion. Although the order of the expansion is not technically
limited, thanks to the systematic calculation
method~\cite{ManoTakasugi}, the highest PN order attainable in practice
is limited by the availability of computational resources.
 In this paper, we revisit the problem of the 
self-force acting on a scalar charge moving in a circular orbit
around a Schwarzschild black hole,
and demonstrate that this practical limitation
of our new method is in fact not at all severe.
We investigate the convergence of the PN expansion
and estimate the \emph{required PN order} to obtain
sufficiently accurate waveforms. We also clearly elucidate the
difference between the roles of the conservative part and the
dissipative part of radiation reaction forces in this formalism.
The role played by the conservative self-force during the orbital evolution
has to this time been demonstrated only in toy model scenarios (see, for
example, Ref.~\citen{Wiseman}). We also discuss the convergence of the
remaining $\tilde R$-part.

The paper is organized as follows. In \S\ref{sec:reg}, we
summarize our new regularization method in the case of the scalar
self-force. In \S\ref{sec:circular}, we calculate the scalar
self-force on a particle in a circular orbit around a
Schwarzschild black hole. First, we explicitly obtain an
expression for the ($\tilde{S}-S$) part of the scalar self-force to 
4 PN order. It is possible to extend the calculation to an arbitrarily high
order systematically. In fact, though not presented explicitly
in this paper, the actual calculation is done to 18PN order.
Then, using this result, we study the convergence of the
($\tilde{S}-S$)-force in the PN expansion.
Subsequently, combining this result with the calculation of the
$\tilde R$-part with sufficient accuracy, so that it does not spoil the
18PN order accuracy of the ($\tilde{S}-S$)-part, the regularized scalar
self-force is evaluated and is compared with the result obtained
by Detweiler, Messaritaki and Whiting~\cite{Detweiler:2002gi}.
In \S\ref{sec:gwcycle}, in order to obtain a rough, qualitative
estimate of the PN order to which we need to proceed in the realistic
gravitational case, we pretend that the scalar charge can be replaced
by the gravitational mass and consider the phase error in the gravitational
waveform due to the truncation of a series in the PN expansion in our
regularization calculation.
The final section, \S\ref{sec:conclusion}, contains conclusion and
discussion on the implications of our result.

\section{Analytic regularization scheme}\label{sec:reg}

We consider a point particle of a scalar charge $q$ moving
in a Schwarzschild background characterized by mass $M$,
\begin{eqnarray}
ds^2 = - \left(1-\frac{2M}{r}\right)dt^2
+\left(1-\frac{2M}{r}\right)^{-1}dr^2
 +r^2 \, (d\theta^2+\sin^2\theta d\phi^2) \, .
\end{eqnarray}
The equation for the scalar field induced by the particle
is given by
\begin{eqnarray}
\nabla^\alpha\nabla_\alpha \psi(x) &=& -q\int_{-\infty}^\infty
d\tau\displaystyle{\frac{\delta^{(4)}(x-z(\tau))}{\sqrt{-g}}} \, .
\end{eqnarray}
Here, $\tau$ and $z(\tau)$ are, respectively, the proper time and
trajectory of the particle, and $g$ is
the determinant of the metric.
The solution for the (full) scalar field can be obtained using the
retarded Green function as
\begin{eqnarray}
\psi^{\rm full}(x) &=& q\int_{-\infty}^{\infty} 
d\tau\, G^{\rm full}(x,z(\tau)) \, .
\label{eq:full-g}
\end{eqnarray}
The retarded Green function satisfies the Klein-Gordon equation
\begin{eqnarray}
\nabla^\alpha \nabla_\alpha G^{\rm full}(x,x') &=&
- \displaystyle{\frac{\delta^{(4)}(x-x')}{\sqrt{-g}}} \,, \label{eq:wave}
\end{eqnarray}
with retarded boundary conditions. Furthermore, due to the 
spherical symmetry and the static nature of the background spacetime,
the Green function can be decomposed in Fourier-spherical
harmonics as
\begin{eqnarray}
G^{\rm full}(x,x') &=& \int \frac{d\omega}{2\pi} \,
e^{-i\omega(t-t')} \sum_{\ell m} g^{\rm full}_{\ell m\omega}(r,r')
Y_{\ell m}(\theta,\phi)Y^*_{\ell m} (\theta',\phi')\,.
\label{eq:FHdeco-full-g}
\end{eqnarray}
Here the functions $Y_{\ell m}(\theta,\phi)$ are the ordinary spherical
harmonics. Then, Eq.~(\ref{eq:wave}) reduces to an ordinary differential
equation for the radial part of the retarded Green function,
\begin{equation}
\left[\left(1-\frac{2M}{r}\right)\frac{d^2}{dr^2}
+\frac{2(r-M)}{r^2}\frac{d}{dr} +\left(\frac{r \omega^2}{r-{2M}}
-\frac{\ell(\ell+1)}{r^2}\right) \right] g^{\rm full}_{\ell
m\omega}(r,r') = -\frac{\delta(r-r')}{r^2}  \,.
\label{eq:radial-eq}
\end{equation}
The radial part of the full Green function is expressed in
terms of two independent homogeneous solutions of Eq.~(\ref{eq:radial-eq})
as
\begin{equation}
g_{\ell m\omega}^{\rm full}(r,r') = \frac{-1}{W_{\ell m\omega}
(\phi_{\rm in}^{\nu},\phi_{\rm up}^{\nu})} \left(\phi_{\rm
in}^{\nu}(r)\phi_{\rm up}^{\nu}(r')\theta(r'-r) +\phi_{\rm
up}^{\nu}(r)\phi_{\rm in}^{\nu}(r')\theta(r-r')\right) \,,
\end{equation}
with $W_{\ell m\omega}$ being the Wronskian,
\begin{eqnarray}
W_{\ell m\omega}(\phi_{\rm in}^{\nu},\phi_{\rm up}^{\nu}) =
r^2\left(1-\frac{2M}{r}\right) \left[ \biggl(\frac{d}{dr}\phi_{\rm
up}^{\nu}(r)\biggr)\phi_{\rm in}^{\nu}(r)
-\biggl(\frac{d}{dr}\phi_{\rm in }^{\nu}(r)\biggr)\phi_{\rm
up}^{\nu}(r) \right] \,.\nonumber
\end{eqnarray}
Here $\phi_{\rm in}^{\nu}$ is a homogeneous solution with
vanishing flux from the past horizon (when multiplied by
$e^{-i\omega t}$), and $\phi_{\rm up}^{\nu}$ is a homogeneous
solution with vanishing flux from the past null infinity.
They can be obtained using a systematic method developed by
Mano, Suzuki and Takasugi~\cite{ManoTakasugi}. The parameter
$\nu$ is called the renormalized angular momentum, which reduces
to $\ell$ in the limit $\omega M \to0$~\cite{ManoTakasugi,ManoTak}.

We can express the homogeneous solutions $\phi_{\rm in}^\nu$ and
$\phi_{\rm up}^\nu$ in terms of another set of solutions,
$\phi_{\rm c}^{\nu}$ and $\phi_{\rm c}^{-\nu-1}$, 
which are given by a series of Coulomb wave
functions as~\cite{ManoTakasugi,Sasaki-Tagoshi}
\begin{eqnarray}
\phi_{\rm in}^{\nu} &=& \phi_{\rm c}^{\nu}
+\tilde{\beta}_{\nu}\,\phi_{\rm c}^{-\nu-1} \,,
\nonumber\\
\phi_{\rm up}^{\nu} &=& \tilde{\gamma}_{\nu}\,\phi_{\rm c}^{\nu}
+\phi_{\rm c}^{-\nu-1} \,.
\end{eqnarray}
The properties of the coefficients $\tilde\beta_\nu$ and
$\tilde\gamma_\nu$ are discussed in Ref.~\citen{ManoTakasugi}.
As shown in one of our previous papers~\cite{HJNSST},
 using these homogeneous solutions, the Green
function can be divided into the two parts as
\begin{eqnarray}
&&g_{\ell m\omega}^{\rm full}(r,r')= g_{\ell m\omega}^{\tilde S}(r,r')
+g_{\ell m\omega}^{\tilde R}(r,r') \,,
\end{eqnarray}
where
where
\begin{eqnarray}
g_{\ell m\omega}^{\tilde S}(r,r')\! &=&\! \frac{-1}{W_{\ell
m\omega}(\phi_{\rm c}^{\nu},\phi_{\rm c}^{-\nu-1})} \left[
\phi_{\rm c}^{\nu}(r)\phi_{\rm c}^{-\nu-1}(r')\theta(r'-r)
+\phi_{\rm c}^{-\nu-1}(r)\phi_{\rm c}^{\nu}(r')\theta(r-r')
\right] , \nonumber
\\
g_{\ell m\omega}^{\tilde R}(r,r')\! &=&\!
\frac{-1}{(1-\tilde\beta_\nu\tilde\gamma_\nu) W_{\ell
m\omega}(\phi_{\rm c}^{\nu},\phi_{\rm c}^{-\nu-1})} \Biggl[
\tilde\beta_\nu\tilde\gamma_\nu \left(\phi_{\rm
c}^{\nu}(r)\phi_{\rm c}^{-\nu-1}(r') +\phi_{\rm
c}^{-\nu-1}(r)\phi_{\rm c}^{\nu}(r')\right) \cr &&\qquad
 \qquad\qquad +\tilde\gamma_\nu \phi_{\rm
c}^{\nu}(r)\phi_{\rm c}^{\nu}(r') +\tilde\beta_\nu \phi_{\rm
c}^{-\nu-1}(r)\phi_{\rm c}^{-\nu-1}(r') \Biggr] .
\end{eqnarray}
The $\tilde S$-part of the radial Green function, 
$g_{\ell m\omega}^{\tilde S}$, is symmetric, and it satisfies
the same inhomogeneous equation as the full radial Green function,
hence becomes singular when the sum over the spherical harmonic
indices is taken. 
Contrastingly, the $\tilde R$-part, $g_{\ell m\omega}^{\tilde R}$, 
satisfies the source free equation, and hence remains regular.
It is the $\tilde S$-part that needs to be regularized.
An important fact is that the $\tilde{S}$-part contains
only positive integer powers of $\omega$ when expanded. Therefore
the frequency integral can be analytically carried out to give
the $\tilde S$-part in the time domain easily.
By contrast, it is difficult to obtain the
$\tilde R$-part in the time domain analytically for general orbits,
because it includes terms logarithmic in $\omega$.
The contribution of such a term in the time domain 
cannot be expressed in terms of local quantities at the location of
the particle, but it can be expressed non-locally in terms of the integral
along the past trajectory. Nevertheless, because the $\tilde R$-part 
is regular and the summation over $\ell$ converges, 
it should be possible to evaluate it numerically 
without much difficulty.

The scalar self-force is given by
\begin{eqnarray}
F_{\alpha} = qP_{\alpha}^{\beta}\partial_{\beta} \psi \, ,
\end{eqnarray}
where the projection tensor 
$P_\alpha{}^\beta=\delta_\alpha{}^\beta+u_\alpha u^\beta$ is
applied to it in order to keep the scalar charge constant. The
regularized self-force is given by the $R$-part
of the self-field defined by Detweiler and Whiting~\cite{DW03}.
The $R$-part of the self-force in our new decomposition scheme
now takes the form
\begin{eqnarray}
F_{\alpha}^{R} = F_{\alpha}^{\rm full} -F_{\alpha}^{S} =
(F_{\alpha}^{\tilde S}-F_{\alpha}^{S}) +F_{\alpha}^{\tilde R} \, .
\label{eq:force-decomp}
\end{eqnarray}
Further motivating the above decomposition of the self-force from a
physical point of view, we note that it has been shown that 
the ($\tilde{S}-S$)-part of the self-force is shown to be purely
conservative for generic  orbits~\cite{HJNSST}. In other words, the
radiative (or dissipative) part of the force is solely contained
in the $\tilde R$-part. This property is an additional merit of our
decomposition.

\section{Convergence test of the PN expansion}
 \label{sec:circular}

Because our regularization method relies on the PN expansion,
it is necessary to examine if it converges sufficiently rapidly.
First we investigate the convergence of a PN series expansion
for the $(\tilde S-S)$-part for circular orbits,
as a test of the efficiency of our analytic regularization method.
To obtain an analytical expression for the $(\tilde S-S)$-part
in the time domain for general orbits,
it is necessary to expand it in powers of $\omega$.
This means that we adopt the slow motion approximation,
\begin{eqnarray}
\omega r =O(v) \,,\quad \omega M = O(v^3) \,;\quad v
~\ll~ 1 \,,
\label{eq:slowmotion}
\end{eqnarray}
where $v$ is the velocity of the particle.
A calculation is said to be of $n$-PN order if 
it is accurate up through $O(v^{2n})$.

We next study the convergence of the $\tilde R$-part.
In the circular case, it is easy to evaluate even the $\tilde{R}$-part
analytically in the time domain.  For a circular orbit, the trajectory
$z^{\alpha}(\tau)$ and the four velocity $u^{\alpha}(\tau)$
of the particle can be written as
\begin{eqnarray}
\{z^\alpha(\tau) \}= \left\{u^t \tau ,\, r_0 ,\, \frac{\pi}{2} ,\,
u^\phi \tau \right\} \,;\quad u^t =
\sqrt{\displaystyle{\frac{r_0}{r_0-3M}}} \,, \quad u^\phi ~=~
\frac{1}{r_0}\sqrt{\frac{M}{r_0-3M}} \, .\ \hspace*{0.7cm}
\label{eq:orbit}
\end{eqnarray}
Then the frequency integral is readily done by substituting
$m\Omega$ for $\omega$, where $\Omega=u^\phi/u^t$. 
However, unlike the case of the $(\tilde S-S)$-part, a time-domain
expression for the $\tilde R$-part cannot be easily obtained for
general orbits. For this reason, we do not bother expanding it in
$\omega$ but test the convergence with respect to the summation over
$\ell$.

In these convergence tests, in order to obtain a rough estimate
for the corresponding tests in the gravitational case,
we translate our results in the scalar case to the gravitational case
by identifying $q/\sqrt{G}$ with the mass $\mu$ of the particle.

\subsection{The ($\tilde{S}-S$)-part}\label{sec:tS-S}

The transformation of the $\tilde S$-part into the time domain makes it
possible to subtract the divergent $S$-part analytically. 
If we were to perform this
subtraction numerically, the fraction to be subtracted would become
closer and closer to unity as $\ell$ increases. Apparently, this would
imply a stringent requirement of numerical accuracy.
In this sense, we anticipate a clear advantage of the analytical
subtraction.

As noted above Eq.~(\ref{eq:slowmotion}), it is necessary to
expand the $(\tilde S-S)$-part in powers of $\omega$,
which corresponds to a PN expansion, to obtain 
its expression in the time domain.
Here, however, we would like to emphasize that the PN expansion
should not be taken as a limitation of our general scheme.
Let us explain the reason. The decomposition into $\ell$ modes
is unavoidable, as long as we adopt the mode-sum (mode-by-mode)
regularization, in which the large $\ell$ modes are the cause of the
divergence. Now, once we regularize the $\tilde S$-part by subtracting
off the $S$-part, the result is a convergent series in $\ell$ modes,
which is essentially a PN series. Thus, we cannot completely avoid PN
expansion even if we employ a fully numerical approach.  

When the orbit shrinks and approaches the innermost stable
circular orbit (ISCO), it is believed that
the PN expansion is a poor approximation. This belief comes partly from 
the expected difficulty in carrying out the PN expansion to very high orders,
and partly from the possible worsened convergence of the PN
expansion, particularly for orbits close to the ISCO. 
We demonstrate that this belief is unfounded on both counts.
Specifically, with regard to the first point, we calculate to 18PN order
to show that systematic calculations to very high PN order are indeed
possible.
With regard to the second point, which is the central issue of this paper,
we find that the convergence is rather rapid, at least for almost
circular orbits, and the expansion up to 4PN or 6PN order turns out
to be sufficient
for the $(\tilde S-S)$-part, even for orbits shrinking up to the ISCO.

The force arising from the $({\tilde S}-S)$-part is purely
conservative for generic geodesic orbits. The conservative part of
the self-force influences the orbit by pushing a particle off the
geodesic orbit, and it also appears in source terms for second
order perturbations. Therefore, it clearly affects the waveforms,
but with corrections of higher order in $\mu/M$. Specifically, when
we investigate the orbital evolution, corrections due to
the conservative self-force are, roughly speaking, given by its
ratio to the Newtonian force. In contrast, since the radiative
part is that responsible for the inspiral, errors due to
the truncation of the PN expansion at a finite order are to be
evaluated in comparison with its leading term.
Hence, errors in the conservative self-force have an
additional suppression factor, $\mu/M$, compared with those in the
radiative part. This is partly the reason why a rather low
PN order, such as 4PN, is adequate to obtain
the $(\tilde S-S)$-part with sufficient accuracy.

The components of the self-force from the ($\tilde{S}-S$)-part of
the scalar field,
$F_{\alpha}^{\tilde S-S}= F_{\alpha}^{\tilde S}-F_{\alpha}^{S}$,
have already been obtained for general orbits in Ref.~\citen{HJNSST}.
This part of the self-force is expressed in terms of 
local quantities of the particle, i.e., its position and velocity.
Therefore, all we need is to provide in
these general formulae information specific to the circular nature of
the orbit, Eq.~(\ref{eq:orbit}). Only the $r$-component is
non-vanishing for the ($\tilde{S}-S$)-force, because the $t$- and
$\varphi$-components are directly related to the rates of change of
the energy and angular momentum, and they are purely dissipative
for circular orbits. 
It is given explicitly as
\begin{eqnarray}
F^{\tilde S-S}_r
&=& \frac{q^2}{4\pi r_0^2}
\biggl[
-\frac{73}{133}+\frac{16151}{21014}V^2
+\frac{395567}{106808}V^4
+\left(\frac{1107284037660637}{400151300487120}
 +\frac{7}{64}\pi^2\right)V^6
\cr && \hspace*{1cm} + \left( - \frac
{182118981911377689978271}{8548630707351386171520} + \frac{29\,\pi
^{2}}{1024} \right)\,V^8 \biggr] + O(v^9) \,,
\end{eqnarray}
where $V=\sqrt{M/r_0}=r_0 \Omega$. Here we have expanded
$F^{\tilde S-S}_r$ up to 4PN order.
 However, it is computed up to 18PN order in the convergence
test below. 

\begin{figure}[ht]
\centerline{
\includegraphics[scale=.81]{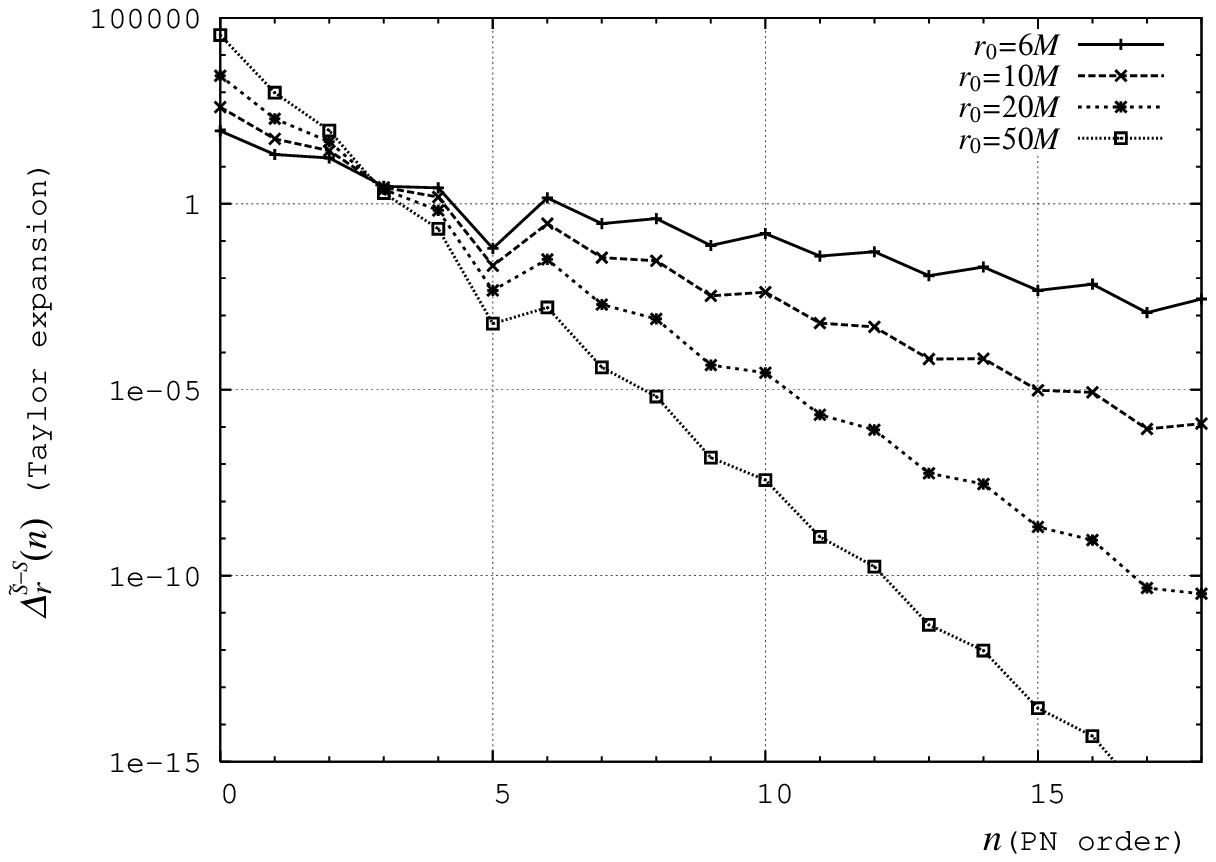}}
\centerline{
 \includegraphics[scale=.8]{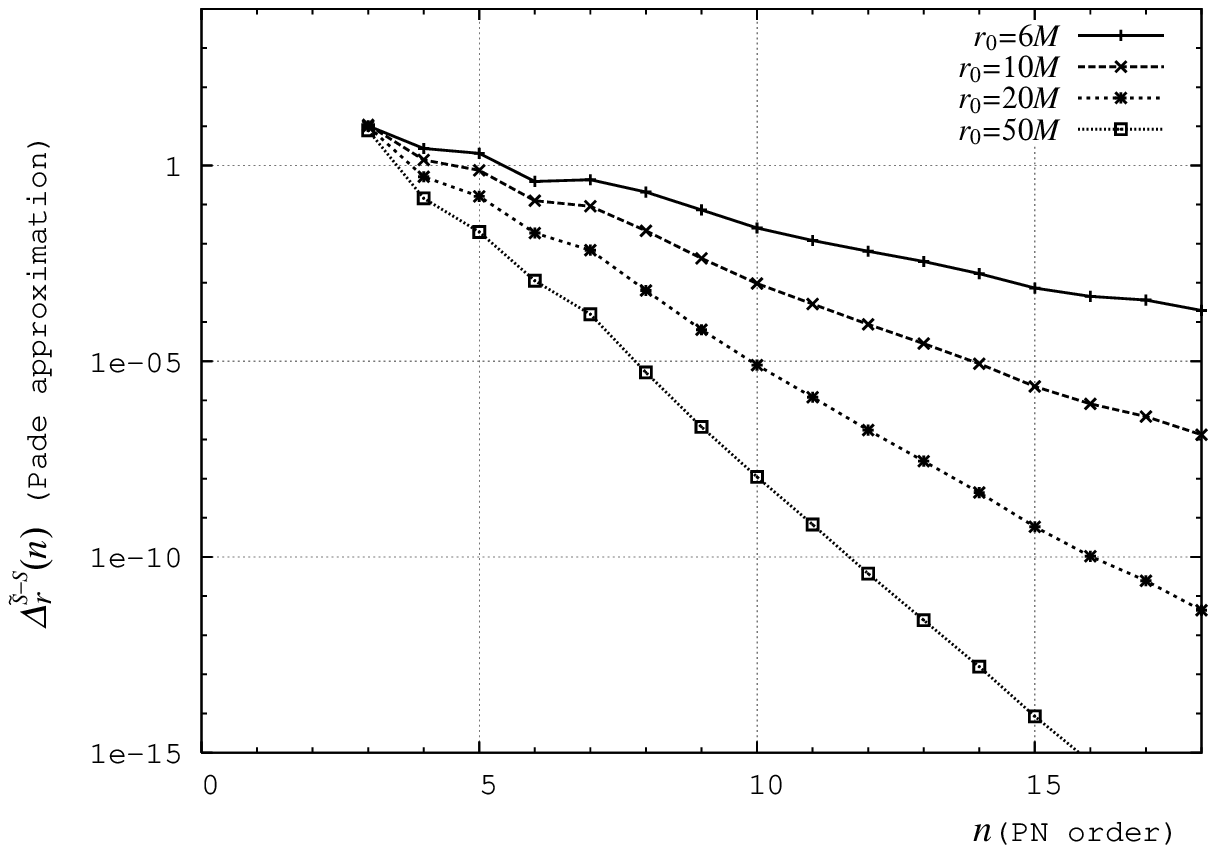}}
         \caption{The relative error of post-Newtonian
formulas in the $r$-component of the ($\tilde{S}-S$)-force 
at $r_0=6M, 10M, 20M$ and $50M$. The horizontal axis is the
order of the post-Newtonian expansion. The top figure displays the
convergence in the Taylor expansion and the bottom figure is that
in the Pade approximation.} \label{fig:FRerr}
\end{figure}

First, in Fig.~\ref{fig:FRerr}, we display the convergence of the
$r$-component of the ($\tilde{S}-S$)-force as a function of the
PN order for several representative orbital radii
$r_0$. Here, an estimator of the convergence of the PN expansion is
 defined by
\begin{equation}
\Delta_{\alpha}^{\tilde S-S}(n)
:= \left|\frac{F_\alpha^{\tilde S-S}|_n -
F_\alpha^{\tilde S-S}|_{n-1}}{F_\alpha}\right|,
\label{conv}
\end{equation}
where $F_\alpha^{\tilde S-S}|_n$ denotes the $(\tilde S-S)$-part of
the force truncated at $n$-PN order, and the
denominator $F_\alpha$ denotes the exact (fully relativistic) self-force
including the $\tilde R$-part. In practice, because it is impossible to
know its exact value, we use the most accurate result in our
calculation. It is found that the
convergence of the PN expansion is steady near the ISCO, although it is
less rapid there. The convergence improves slightly when the Pade
approximation is used near the ISCO. Here, in the Pade approximation, we
have chosen the denominator to be quadratic in $V^2$.

\subsection{The $\tilde{R}$-part and the regularized self-force}
\label{sec:SSF}

In the case of circular orbits around a Schwarzschild black hole,
each component of the self-force due to the $\tilde{R}$-part is
formally given by
\begin{eqnarray}
F_t^{\tilde{R}} &=&
-\frac{iq^2\Omega}{u^t}\sum_{\ell m} m \,
g_{\ell m, m\Omega}^{\tilde{R}}(r_0,r_0)
\left|Y_{\ell m}\left(\frac{\pi}{2},0\right)\right|^2,
\nonumber \\
F_r^{\tilde{R}} &=&
\frac{q^2}{u^t}\sum_{\ell m} \left.
\partial_r g_{\ell m, m\Omega}^{\tilde{R}}(r,r_0)\right|_{r=r_0}
\left|Y_{\ell m}\left(\frac{\pi}{2},0\right)\right|^2,
\nonumber \\
F_\theta^{\tilde{R}} &=& 0,
\nonumber \\
F_\phi^{\tilde{R}} &=&
-\frac{1}{\Omega}F_t^{\tilde{R}}. \label{eq:tilR}
\end{eqnarray}
The $\tilde{R}$-force in this case can be completely
obtained analytically, because the integration with
respect to $\omega$ is done by substituting $m \Omega$ for
$\omega$. Also, note that since only modes up to a finite value of $\ell$
contribute to the self-force for a given PN order, the computation of
the $\tilde{R}$-force is relatively easy. If we need 
precision up to $n$-PN order inclusive, it is sufficient for us to
calculate the modes up to $\ell \le n+1$. The detailed analysis is
summarized in Ref.~\citen{HJNSST}.
 The 4PN results, after summation over the 
$\ell$-modes, are
\begin{eqnarray}
F^{\tilde R}_r &=& {q^2 \over 4\pi r_0^2} \biggl[ {73 \over
133}-{16151 \over 21014}V^2-{395567 \over 106808}V^4 \cr&&
- \left({4 \over 3}\gamma+{4 \over 3}\ln(2V) +{1196206548879997 \over
400151300487120}\right)V^6 \cr &&
+\left(\frac{59372120592232147984979}{1709726141470277234304} -
\frac{14}{3} \,\ln(V) - \frac{66}{5} \,\ln(2) - \frac
{14\,\gamma}{3} \right)V^8  +O(v^9) \biggr]\,, \cr
F^{\tilde R}_t &=& {q^2 V\over 4\pi r_0^2} \biggl[ {1 \over 3}V^3-{1 \over
6}V^5+{2\pi \over 3}V^6 -{77 \over 24}V^7 +{9\pi \over 5}V^8
+ O(v^{9})\biggr] \,.
\end{eqnarray}
Here $\gamma$ is Euler's constant, $\gamma=0.57\cdots$.
The temporal component, which represents the energy loss rate,
starts at $1.5PN$ order, as is expected for dipole radiation.

\begin{figure}[ht]
\begin{center}
\centerline{
\includegraphics[scale=.8]{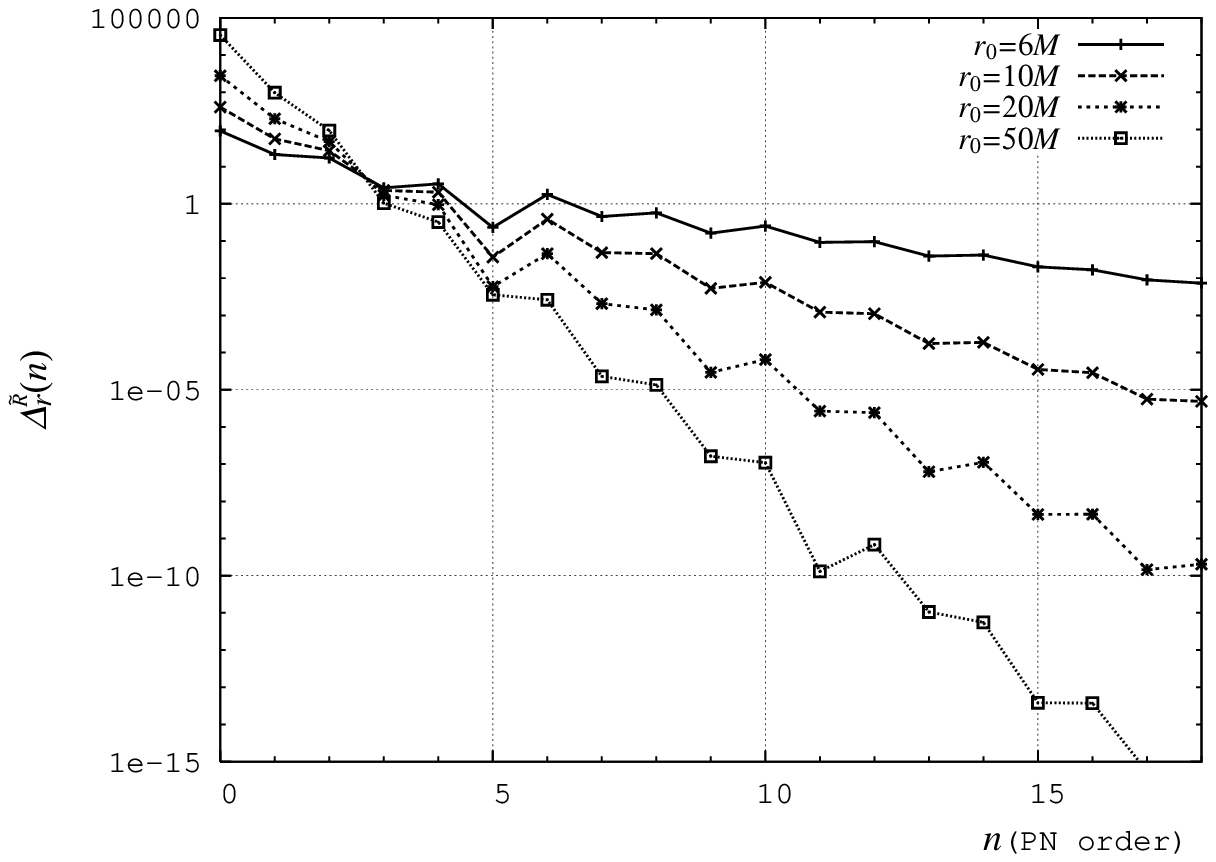}}
\centerline{
\includegraphics[scale=.8]{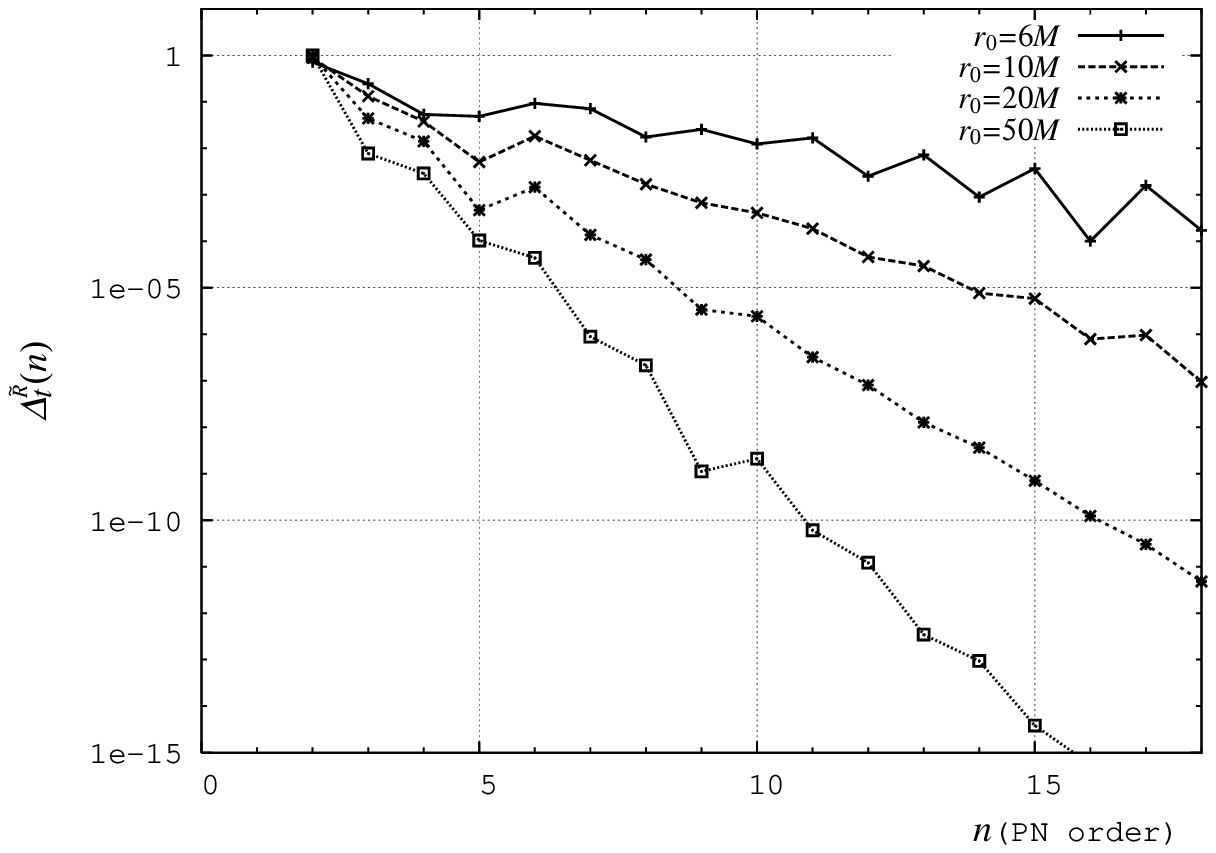}}
\caption{Plots of errors in the post-Newtonian expansion of
both the $r$-component (upper panel) and
the $t$-component (lower panel) of the $\tilde{R}$-force. }
\label{fig:FtilRerrPN}
\end{center}
\end{figure}

\begin{figure}[ht]
\begin{center}
\centerline{
\includegraphics[scale=.8]{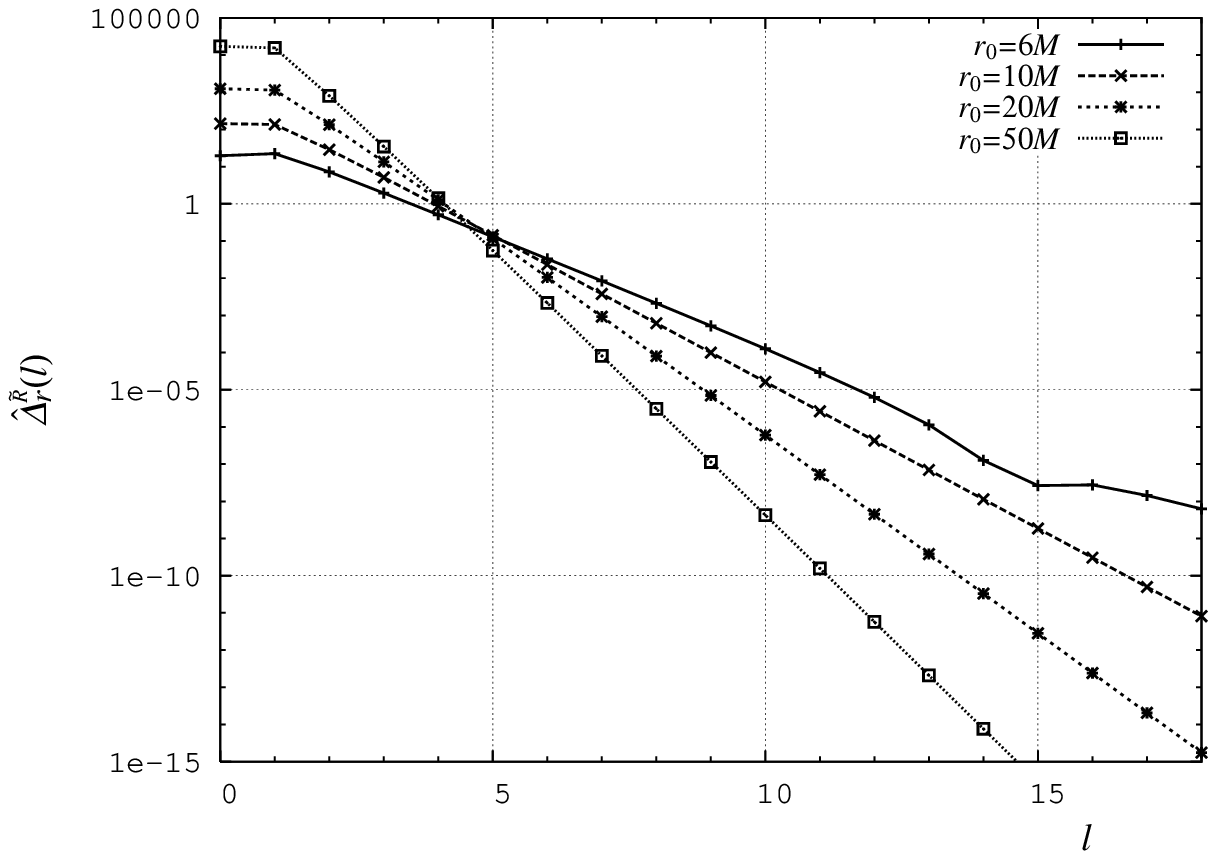}}
\centerline{
\includegraphics[scale=.8]{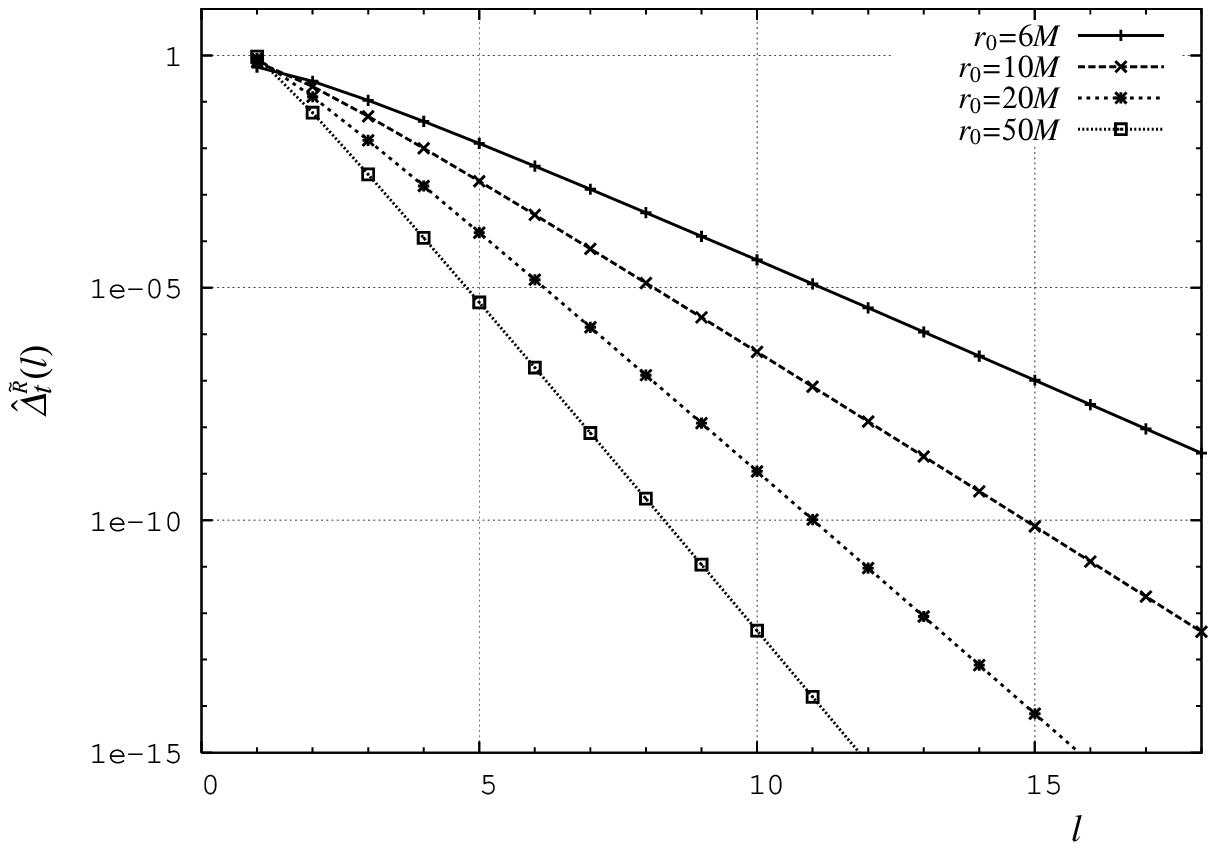}}
         \caption{Plots of the errors in the $\ell$ expansion of both
the $r$-component (upper panel) and the $t$-component (lower panel) of
the $\tilde{R}$-force. Here, $\hat \Delta_{\alpha}^{\tilde R}(\ell)$ is
defined in the same way as in the case of $\Delta_{\alpha}^{\tilde
R}(n)$, but in the $\ell$-expansion instead of the post-Newtonian
expansion. The convergence is much faster than that of the naive
post-Newtonian expansion.} \label{fig:FtilRerr}
\end{center}
\end{figure}

To see the convergence of both
the $r$-component and $t$-component of the $\tilde R$ force, in
Fig.~\ref{fig:FtilRerrPN} we plot $ \Delta_{\alpha}^{\tilde R}(n)$,
defined in a manner analogous to $\Delta_{\alpha}^{\tilde S-S}(n)$,
for several representative orbital radii $r_0$. The
$\tilde R$-part contains the radiative part, whose errors are to be
compared with its leading term, as mentioned above. Thus we need
to evaluate the $\tilde R$-part with better accuracy than the
$(\tilde S-S)$-part. In this case, it would seem that the convergence
of the naive post-Newtonian expansion given above is too slow for a
small $r_0$. Since the
$\tilde R$-part plays no role in the divergences, it is not
necessary to transform its expression into the time domain
analytically. This means that we do not have to expand the 
$\tilde R$-part of the Green function in powers of $\omega$.
Here we propose to use just the spherical harmonic expansion
($\ell$-expansion), and to compute the contribution from each
$\ell$-mode with sufficient accuracy. A criterion
for truncating various series expansions in 
$g_{\ell m\omega}^{\tilde R}(r,r')$ is discussed in the Appendix in some
detail. Plots of the $\ell$-expansion similar to that in 
Fig.~\ref{fig:FtilRerrPN} are given in Fig.~\ref{fig:FtilRerr}.
\newpage
\subsection{Comparison with Detweiler, Messaritaki and
 Whiting~\cite{Detweiler:2002gi}}

The total self-force in the case of a circular orbit around a
Schwarzschild black hole is obtained by summing the
($\tilde{S}-S$)-part and the $\tilde{R}$-part as
\begin{eqnarray}
F^{R}_r &=& {q^2 \over 4\pi r_0^2} \biggl[ \left( -{4 \over
3}\gamma+{7 \over 64}\pi^2-{4 \over 3}\ln(2V) -{2 \over
9}\right)V^6 \cr && +\left( {604 \over 45} + {29\pi^2 \over 1024}
- {66 \over 5}\ln(2) - {14 \over 3}\ln(V) - {14 \over 3}\gamma
\right)V^8
\biggr] + O(v^9) \,, \label{eq:FRr_cir}\\
F^{R}_t &=& F^{\tilde R}_t.
\end{eqnarray}
In the $r$-component of the scalar self-force, there is a
significant cancellation between the $(\tilde S-S)$-part and the
$\tilde R$-part, and the total force begins at 3PN order.

Here we compare our result with that obtained
by Detweiler, Messaritaki and Whiting~\cite{Detweiler:2002gi}.
They calculated the radial component of the self-force for the
case $r_0=10M$, $M=1$ and $q^2=4\pi$, and evaluated the
uncertainty using a Monte Carlo simulation. It should be noted
that the source term of the field equation in our definition is
different from theirs by a factor of $4\pi$. Their
result is
\begin{eqnarray*}
F^{R}_r = 1.37844828(2)\times 10^{-5} \, .
\end{eqnarray*}
On the other hand, with the result obtained using a Pade approximation
for the ($\tilde{S}-S$)-force accurate to 18PN order and with the most
accurate $\tilde{R}$-force in our calculation including the terms
up to $\ell=18$, we obtain
\begin{eqnarray*}
F^{R}_r = 1.378448171\times 10^{-5} \, .
\end{eqnarray*}
In Table~\ref{tab:FR}, we list the numerical values for the
$r$-component of the regularized self-force for $r_0=6M$ (ISCO),
$10M$, $20M$ and $r_0=50M$. In our computation the
accuracy is limited by the ($\tilde{S}-S$)-part. Hence, the
accuracy of the full regularized force can be read from
Fig.~\ref{fig:FRerr}.

\begin{table}[t]
\caption{The $r$-component of the self-force at several
radii ($M=1$ and $q^2=4\pi$). }\label{tab:FR}
\begin{center}
\setlength{\tabcolsep}{1mm}
\begin{tabular}{c|c|c|c|c}
 \hline\hline
 $r_0$ & $6M$ & $10M$ & $20M$ & $50M$\\
\hline
 $F_r^{{\rm R}}(r_0)$
   &$1.676820878\times10^{-4}$ &$1.378448171\times10^{-5}$
   &$4.937905866\times10^{-7}$ &$6.346791373\times10^{-9}$ \\
 \hline
\end{tabular}
\end{center}
\end{table}

\section{Errors in gravitational wave cycles}
\label{sec:gwcycle}

In this section, we evaluate the PN order necessary
to obtain sufficiently accurate gravitational waveforms. For this
purpose, we consider the number of wave cycles $N$ when the
particle spirals in from an initial radius $r_i$ to a final radius
$r_f$.

For the gravitational wave search with known waveforms, the
so-called {\it matched filtering method} is used. In this
method, the correlation between the detector's output and theoretical
templates for the waveforms is taken to search for the maximum
signal-to-noise ratio (SNR). Because a phase error of order unity
in the theoretical templates
strongly reduces the SNR, an error in the number of cycles, denoted by 
$\Delta N$, caused by truncating the force at a given order of the
expansion is a good indicator of the significance of the error.
Here, because the frequency of the gravitational wave is given by twice
the orbital period, we estimate twice the number of cycles of the
orbital rotation by the formula
\begin{eqnarray}
N &=&
\frac{1}{\pi}\int_{r_i}^{r_f}\Omega\frac{(dE/dr_0)}{dE/dt}dr_0
 \cr
 & \approx & {4M\over\mu}\int_{V_i}^{V_f}dV
    {(1-6V^2)\over (1-3V^2)^2f_t^R},
\label{Nformula}
\end{eqnarray}
where $V_i=\sqrt{M/r_i}$, $V_f=\sqrt{M/r_f}$, and 
$f_t^R=(4\pi r_0^2/q^2)\times F_t^R$.

First, we consider the correction due to the conservative part of the
self-force. The factor $\Omega (dE/dr_0)$ in Eq.~(\ref{Nformula})
is determined by the background geodesic motion at lowest
order. Hence, the self-force contribution is suppressed by its
ratio to the Newtonian force. This ratio is given in the PN expansion
as
\begin{eqnarray}\label{cons-force-error}
 {F_r^{\tilde{S}-S}\over (\mu M/r^2)}
  =-\frac{\mu}{M}\sum_{i}a_i V^{2i}.
\end{eqnarray}
Hence, the correction to $N$ due to the $(\tilde S-S)$-force
at $n$-PN order is given by
\begin{eqnarray}
\label{error-cycle-cons}
\Delta N^{\tilde S-S}(n)
 = 4a_n\int_{V_i}^{V_f}\
dV\frac{(1-6V^2)}{(1-3V^2)^2f_t^R}V^{2n}.
\end{eqnarray}
Because the ratio $F_r^{\tilde{S}-S}/(\mu M/r^2)$ contains a factor
$\mu/M$, the $\mu$-dependence in this correction appears only
through $V_i$ and $V_f$. We set $V_f$ equal to the velocity at the ISCO,
where higher PN order corrections become the largest. Explicitly,
$V_f$ is fixed to $1/\sqrt{6}$. The value of $V_i$ depends on the masses
$\mu$ and $M$ and the observation period, but here we set $V_i$ equal to
$0$ in order to estimate the maximum error. This makes 
$\Delta N^{\tilde S-S}(n)$ completely independent of $\mu$.
 Table~\ref{tab:cons-cycle} lists the error estimates.
From this table, it is evident that the expansion up to 4PN order
is in most cases sufficient for the
$(\tilde{S}-S)$-part of the self-force. However, in the case we expect to
detect an inspiral signal with a very large SNR, the observational error
in the number of cycles, $\Delta N$, may be made small,
say, $\Delta N\lesssim 0.1$. In such cases,
in order to extract as much orbital information as possible,
it will be necessary to calculate up to 6PN order or higher.

 \begin{table}[b]
  \caption{ The relative error $\Delta N^{\tilde S-S}(n)$
for the conservative self-force.}
\label{tab:cons-cycle}
\begin{center}
  \begin{tabular}{c||c|c|c|c|c|c|c|c}
   \hline\hline
   $n$ &$2$ &$3$ & $4$ & $5$ & $6$ & $7$ & $8$ & $9$ \\ \hline
   $\Delta N_n$ & $15.1$ & $ 0.660$ & $ 0.292$
& $ 0.00413$ & $ 0.0635$ & $ 0.00927$
& $ 0.00962$ & $ 0.00143$ \\
   \hline
  \end{tabular}
\end{center}
 \end{table}

Next we consider the dissipative part of the self-force,
which is completely contained in the $\tilde R$-part.
In particular, in the case of circular orbits, 
the $t$- and $\varphi$-components of the $\tilde R$-force
are purely dissipative.
Because the self-force is responsible for the orbital energy loss,
the energy loss rate $dE/dt$ which appears in Eq.~(\ref{Nformula})
is solely determined by $F_t^R$.
Therefore, in contrast to the conservative part, the
higher PN corrections of the dissipative self-force are not suppressed
by a factor of $\mu/M$ in comparison with its leading order term.
Hence, the PN corrections through $dE/dt$ is not suppressed by a factor
of $\mu/M$.
We therefore define the error indicator $\Delta N^{\tilde R}(\ell)$ by
\begin{eqnarray}
\label{gen-error-cycle}
\Delta N^{\tilde R}(\ell) &=&
N^{\tilde R}(\ell) - N^{\tilde R}(\ell-1),
\end{eqnarray}
with
\begin{eqnarray}
N^{\tilde R}(\ell) &=& 4\frac{M}{\mu} \int_{V_i}^{V_f}dV
\frac{(1-6V^2)}{(1-3V^2)^2f_t^{R(\ell)}},
\end{eqnarray}
where $f_t^{R(\ell)}$ is the $\tilde R$-part of the force
summed up to the $\ell$-th harmonics.

Interestingly, when both the mass of an inspiraling small compact
object and the observation period are fixed, the error 
$\Delta N^{\tilde R}(\ell)$ has a maximum as a function of the mass of
the central black hole. Hence, an upper bound on the required order of
the $\ell$-expansion can be obtained by evaluating the phase error
formula (\ref{gen-error-cycle}) at the maximum. The presence of
a maximum can be understood by considering two limiting cases:

\begin{itemize}

\item \underline{Small $M$ limit:}. The largest truncation error
comes at the ISCO. Therefore, we fix $V_f$ at the ISCO. Then $V_i$ is
fixed for a given observation period. Here, we adopt $1$ year for
the observation period. When $M$ is small, $V_i$ is small. Thus we
can obtain a good upper bound on the error in $\Delta N$ by setting
$V_i=0$. After substituting $0$ for $V_i$, the
expression~(\ref{gen-error-cycle}) for $\Delta N$
is manifestly proportional to $M$.
Hence, $\Delta N$ decreases as $M$ decreases.

\item \underline{Large $M$ limit:}. Again, we fix $V_f$ at the ISCO.
Then, because $M$ is large, the orbit remains close to the ISCO.
In the limiting case, we can regard the integrand
in Eq.~(\ref{gen-error-cycle}) as constant. Then, we have
$\Delta N \approx N \Delta F_t/F_t$, and the $M$ dependence appears only
through $N$. Because for a given observation period, $N$ decreases
as $M$ increases, $\Delta N$ also decreases as $M$ increases.
\end{itemize}

\begin{figure}[ht]
\begin{center}
\includegraphics[scale=.8]{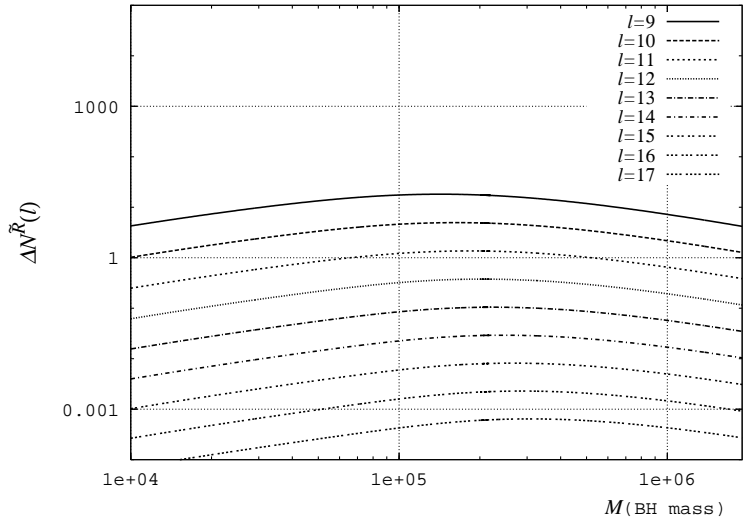}
\caption{ Error in the number of cycles caused by truncating the
$\ell$-expansion in the $\tilde R$-force as a function of the 
mass of the central black hole.} \label{fig:rot_error}
\end{center}
\end{figure}

\noindent
Because $\Delta N$ decreases for both small and large $M$
limits, it should have a maximum. Therefore, we can identify the
``required'' order of the $\ell$-expansion for a given observation
period independently of $M$. We plot $\Delta N^{\tilde R}(\ell)$
 as a function of $M$ in Fig.~\ref{fig:rot_error}, where
we set $\mu$ equal to $1\,M_\odot$ and the observation period to 1 year.
The plots exhibit a peak near $M=10^5M_\odot$,
and the error decreases as $M$ moves away from this value on both sides.
{}From these plots, for an inspiraling compact object of solar mass
and for a given observation period of $O(1{\rm year})$,
the expansion up to $\ell=13$ is found to be sufficient irrespective
of the mass of the central supermassive black hole.

\section{Conclusion and discussion}\label{sec:conclusion}

The new regularization method proposed in Ref.~\citen{HJNSST} is
based on the post-Newtonian (PN) expansion.
To exhibit the effectiveness of our
regularization method, we have considered the simple case of
a scalar charged particle in circular orbits around a Schwarzschild
black hole, and we have shown here that one can actually
compute the PN expansion up to sufficiently high orders.

We have analytically obtained the $(\tilde S-S)$-part of the
scalar self-force up to 18PN order (though the explicit results
are displayed only up to 4PN). Using this result, we investigated
the convergence of the PN expansion for the ($\tilde{S}-S$)-part.
The results are shown in Fig.~\ref{fig:FRerr}. It appears that
our result up to 18PN realizes an accuracy of $\sim$ $10^{-7}$ for
$r_0=10M$.
At the innermost stable circular orbit (ISCO), the convergence slows
down, but is found to be steady, and the accuracy
of the obtained regularized self-force is high, with relative
error of $O(10^{-4})$.

For the $\tilde R$-part, we have computed the contribution from each
spherical harmonic mode up to $\ell=18$. The expansions with
respect to $\epsilon=2M\omega$ and $U=M/r_0$ are truncated at a
sufficiently high order. Because the $\tilde R$-part is unaffected by 
the regularization procedure, we do not have to perform its PN
expansion. Interestingly, the convergence of the $\ell$-expansion
is found to be much faster than the PN expansion, although the results
truncated at the $\ell$-th harmonics are correct only up to
$(\ell+1)$-PN order.

Next, we have considered the phase error in the gravitational 
waveform due to truncation of the PN expanded series
by interpreting the scalar charge as the mass of an orbiting
particle. 
We have found that the 4PN order calculation of the $(\tilde S-S)$-part
seems to be sufficiently accurate. For
the $\tilde{R}$-part, we need a calculation up to $\ell=13$ to make
templates for a one-year observation up to the ISCO, assuming that the
mass of the inspiraling star is 1\,$M_\odot$.

Comments are now in order concerning the convergence behavior
considered in the Appendix. First, we note here that the expression for
the Green function given in Ref.~\citen{ManoTakasugi} is convergent for
any values of $\epsilon$ and $z$. However, the convergence is not
guaranteed once we expand the Green function in powers of these
parameters. In fact, the expansion of the $(\tilde S-S)$-part of the
Green function with respect to $\epsilon$ has a finite convergence
radius. This can be understood as follows\cite{disTagoshi}.
In the limit $\epsilon\to0$, we have $\nu\to\ell$.
Now, as the value of $\epsilon$ increases, $\nu$ decreases and
eventually becomes $\ell-1/2$. For this value of $\nu$,
$\phi^{\nu}$ and $\phi^{-\nu-1}$ are no longer independent,
and the Wronskian $W(\phi^\nu,\phi^{-\nu-1})$ vanishes there.
Therefore, we have a simple pole in $g^{\tilde S}_{\ell m\omega}(r,r')$,
and hence the power series expansion fails to converge at this value
of $\epsilon$. This problem is hidden as far as circular orbits
are concerned, because high frequency contributions are completely
suppressed. However, for generic orbits, we have arbitrarily high
frequency contributions. For example if we expand the radial component
of an orbit with respect to the eccentricity $e$,
terms of higher order in $e$ have higher frequencies.
Hence, $\omega$ can be arbitrarily large.
These high frequency contributions, which cannot be
handled by our analytic regularization method, are present in general.
Nevertheless, we believe that our scheme is useful in a relatively
wide region of the orbital parameter space.

In addition to the problem mentioned above, the spins of the central
black hole and the inspiraling compact star also shift the orbital
frequency at the ISCO. We may encounter situations in which relativistic effects
become more important than we have considered in this paper.
Therefore, developing an alternative numerical method~\cite{Ori:2000zn}
that can play a complimentary role is also important.

Ultimately, our goal is to derive the gravitational
self-force on a point particle orbiting a ``Kerr'' black hole for a
generic orbit, and eventually to construct highly precise theoretical
template waveforms, to be used in the upcoming era of 
gravitational wave astronomy.
 For this purpose, we need to develop the second-order
black hole perturbation theory. Our present results suggest
that a relatively low PN order is sufficient for the second-order
perturbation.
This fact encourages us to develop a formalism along the lines
of our scheme whose treatment of the singular part exploits the PN
expansion efficaciously.

\section*{Acknowledgements}

We would like to thank R. Fujita, H. Ishihara, Y. Mino, T.
Nakamura, K. Nakao and H. Tagoshi for useful discussions. We also
thank all the participants in the Sixth Capra meeting and the Post
Capra meeting at Yukawa Institute, Kyoto University
(YITP-W-03-02). We also thank all the participants in the Seventh
Capra meeting held at the Center for Gravitational Wave Astronomy,
University of Texas at Brownsville. HN is supported by a
JSPS Research Fellowship for Young Scientists, No.~5919. SJ
acknowledges support under a Basque government postdoctoral fellowship.
This work was supported in part by Monbukagaku-sho Grants-in-Aid
for Scientific Research, Nos.~14047212, 14047214 and 12640269, by
the Center for Gravitational Wave Physics, PSU, which is supported by
the NSF under Cooperative Agreement PHY~0114375, and by the Center for
Diversity and Universality in Physics at Kyoto University,
which is supported by the 21st Century COE Program of
Monbukagaku-sho.

\appendix

\section*{Criteria for Truncating in $g^{\tilde{R}}_{\ell m\omega}$}

Because the infinite sum over the spherical harmonic modes
is practically impossible,
we need to decide at which value of $\ell$ to truncate the sum,
which should be determined by an accuracy requirement.

\begin{figure}[t]
\begin{center}
\includegraphics[scale=.7]{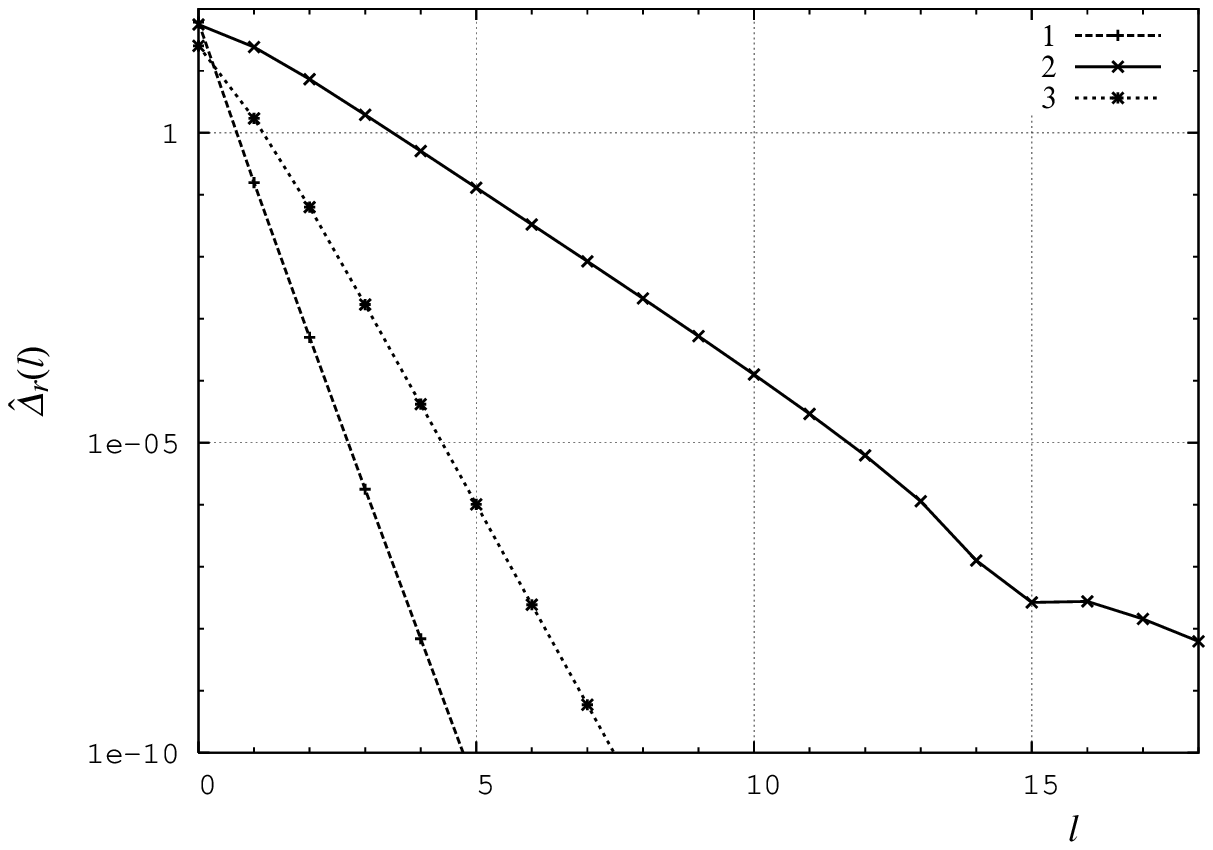}
\includegraphics[scale=.7]{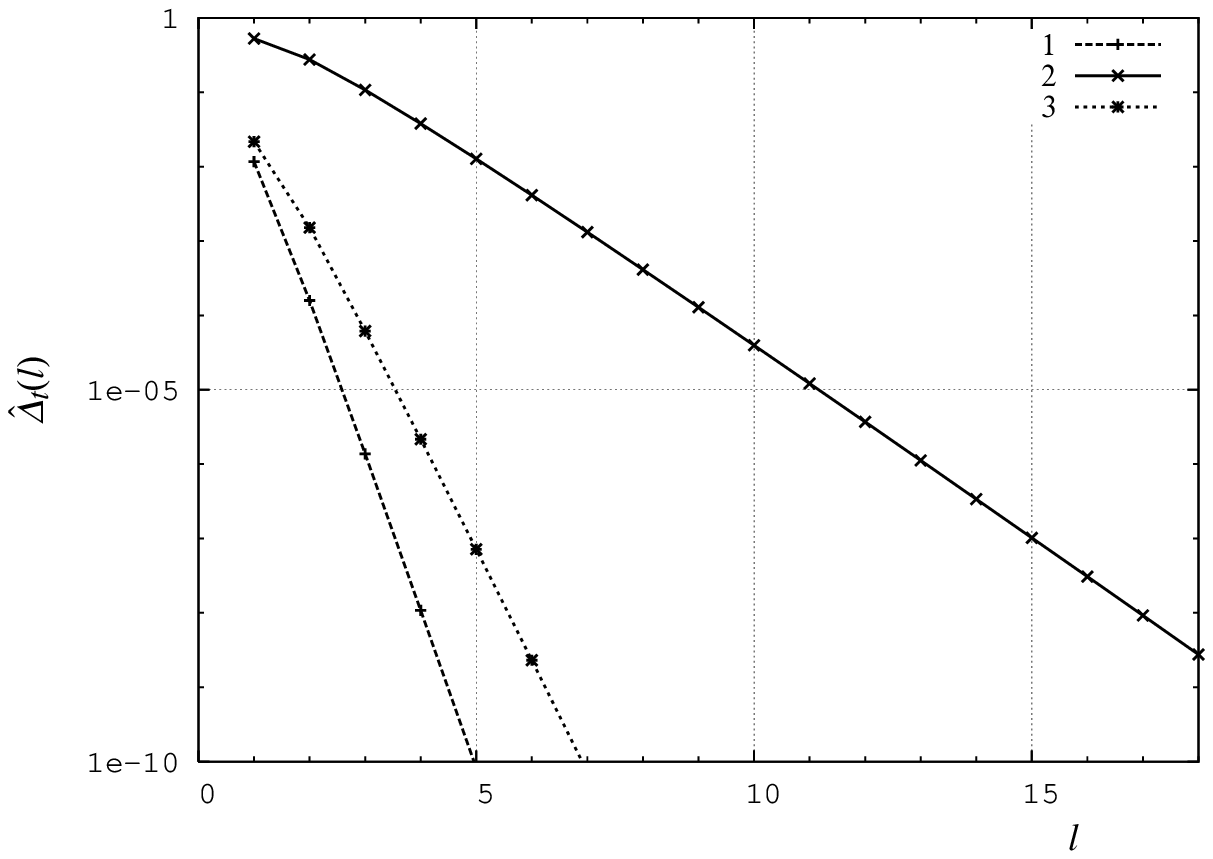}
\caption{Plots of the error in the $\ell$-expansion of both
the $r$-component (upper panel) and the $t$-component (lower panel) of
$F^{1,(8,25)}$, $F^{2,(8,25)}$ and  $F^{3,(8,25)}$ 
(denoted by $1$, $2$ and $3$, respectively) at $r_0=6M$.}
\label{appfig:FtilRerr}
\end{center}
\end{figure}

For the $t$-component, if we compute the sum up to $\ell=18$,
the accuracy is $\sim 10^{-8}$ for $r_0=6M$,
as shown in Fig.~\ref{fig:FtilRerr} or Fig.~\ref{appfig:FtilRerr}.  
Approximately the same accuracy is obtained by the
truncation at $\ell=18$ for the $r$-component as well.
However, because an accuracy of the $r$-component is
determined by that of the ($\tilde{S}-S$)-part,
if we require an accuracy of $\sim 10^{-4}$ for $F^{\tilde{S}-S}_r$,
we only need to compute up to $\ell=11$ for $F^{\tilde{R}}$.

The above conclusion, however, assumes that $g^{\tilde{R}}_{\ell m\omega}$
for each given $\ell$ is computed with sufficiently high accuracy.
In this appendix, we explain our method for computing
 $g^{\tilde{R}}_{\ell m\omega}$ and examine its accuracy.
Our method is based essentially on the post-Minkowski expansion, in
which the black hole mass $M$ plays the role of the expansion parameter.
Specifically, we expand the $\tilde{R}$-part of the Green
function with respect to $\epsilon=2M\omega=O(v^3)$ and
$U=M/r=O(v^2)$.

\subsection{Analytic expressions for the $\tilde R$-part}
The $\tilde{R}$-part of the Green function is given by
\begin{eqnarray}
g_{\ell m\omega}^{\tilde R}(r,r') &=&
\frac{-1}{(1-\tilde\beta_\nu\tilde\gamma_\nu) W_{\ell
m\omega}(\phi_{\rm c}^{\nu},\phi_{\rm c}^{-\nu-1})} \Biggl[
\tilde\beta_\nu\tilde\gamma_\nu \left(\phi_{\rm
c}^{\nu}(r)\phi_{\rm c}^{-\nu-1}(r') +\phi_{\rm
c}^{-\nu-1}(r)\phi_{\rm c}^{\nu}(r')\right) \cr &&\qquad
 \qquad\qquad +\tilde\gamma_\nu \phi_{\rm
c}^{\nu}(r)\phi_{\rm c}^{\nu}(r') +\tilde\beta_\nu \phi_{\rm
c}^{-\nu-1}(r)\phi_{\rm c}^{-\nu-1}(r') \Biggr] \,.
\end{eqnarray}
Using the variables $z=\omega r$ and $\epsilon=2M\omega$,
the function $\phi_{\rm c}^{\nu}$ and the coefficients $\tilde{\beta}$
and $\tilde{\gamma}$ are given by
\begin{eqnarray}
\phi_{\rm c}^\nu (z) &=& e^{-i(z-\epsilon)}
\left(2(z-\epsilon)\right)^{\nu}
\sum_{n=-\infty}^{\infty} i^n a_n{}^\nu
{\Gamma(n+\nu+1+i\epsilon) \over \Gamma(\nu+1+i\epsilon)}
{\Gamma(2\nu+2) \over \Gamma(2n+2\nu+2)}
\left(2(z-\epsilon)\right)^{n}
\cr && \times
{}_1 F_1(n+\nu+1+i\epsilon;\,2n+2\nu+2;\,2i(z-\epsilon)) \,, \label{eq:pc}
\\
\tilde\beta_{\nu} &=&
{\Gamma(-\nu+i\epsilon) \over \Gamma(\nu+1+i\epsilon)}
{\Gamma(2\nu+2) \over \Gamma(-2\nu)}
{K_{-\nu-1} \over K_{\nu}} \,,
\\
\tilde\gamma_{\nu} &=&
i \,{\Gamma(\nu+1+i\epsilon) \over \Gamma(-\nu+i\epsilon)}
{\Gamma(-2\nu) \over \Gamma(2\nu+2)}
e^{-i\pi \nu}{\sin\pi(\nu+i\epsilon)\over \sin\pi(\nu-i\epsilon)}
\,,
\end{eqnarray}
where ${}_1 F_1$ is the confluent hypergeometric function, and
\begin{eqnarray}
K_\nu &=& {2^{-\nu} \epsilon^{-\nu} \Gamma(1-2i\epsilon)
\over
\Gamma(1+\nu+i\epsilon)^2\Gamma(1+\nu-i\epsilon)}
\sum_{n=0}^\infty {\Gamma(n+\nu+1+i\epsilon)\Gamma(n+2\nu+1)
\over
n! \Gamma(n+\nu+1-i\epsilon)}a_n{}^\nu
\cr && \times
\left[\sum_{n=-\infty}^0
{1 \over
(-n)! \Gamma(n+2\nu+2)}
a_n{}^\nu \right]^{-1} \,. \nonumber
\end{eqnarray}
We assume $\omega >0$ in the above derivation of the
homogeneous solutions.
To obtain the solutions for $\omega <0$,
we consider their analytic continuation to complex $\omega$
through the upper half complex $\omega$-plane. This is done by
replacing $\omega$ in $\phi_c^{\nu}$ with $|\omega|e^{i\pi}$.

The Wronskian is given by
\begin{eqnarray}
W_{\ell m\omega}(\phi_{\rm c}^{\nu},\phi_{\rm c}^{-\nu-1})
&=& 
-\frac{2\nu+1}{2\omega}
\left| \sum_{n=-\infty}^{\infty} a_{n}^{\nu}\right|^2.
\end{eqnarray}

It is important to note that we can easily obtain expressions for
 $a_{n}^\nu$ (for $-N<n<N$ where $N$ is a large integer) to 
very high accuracy in powers of $\epsilon$, for any value of $\nu$.
In fact, because the convergence condition of the quantities 
$a_n^\nu$ for $n\to\pm\infty$ determines $\nu$, it is the accuracy of
$\nu$ that is important. Therefore, the accuracies of the coefficients
$\tilde{\beta}$, $\tilde{\gamma}$ and
$W_{\ell m\omega}(\phi_{\rm c}^{\nu},\phi_{\rm c}^{-\nu-1})$ 
essentially depend only on $\nu$.
In other words, we may regard these as functions of only $\nu$
when we estimate their accuracies.

\subsection{Error analysis}
Apart from the dependence on $\nu$, which we leave unexpanded
in all the quantities, we perform the post-Minkowski expansion
by replacing $z$ with $\epsilon/(2U)$. We then assume $\epsilon=O(v^3)$
and $U=O(v^2)$.
We express the homogeneous solution $\phi^{\nu}_c$ as
\begin{eqnarray}
 \phi^\nu_{c}(r)=\epsilon^\nu\left(\frac{1}{U}-2\right)^\nu
\hat{\Phi}^{\nu}_{c}(r).
\end{eqnarray}
We then write a truncated expression for $g^{\tilde{R}}_{\ell
m\omega}$ as
\begin{eqnarray}
 g^{\tilde{R},(n_1,n_2)}_{\ell m\omega}(r,r') =
\G_{\ell m\omega}^{1,(n_1,n_2)}(r,r')+\G_{\ell m\omega}^{2,(n_1,n_2)}(r,r')
+\G_{\ell m\omega}^{3,(n_1,n_2)}(r,r')\,,
\end{eqnarray}
where
\begin{eqnarray}
 \G_{\ell m\omega}^{1,(n_1,n_2)}(r,r')&\equiv&
A_{1}(\nu)
\left[\hat{\Phi}^{\nu}_{c}(r)\hat{\Phi}^{-\nu-1}_c(r')\right]^{(n_1,n_2)}
+(r\leftrightarrow r')
,\cr
\G_{\ell m\omega}^{2,(n_1,n_2)}(r,r')&\equiv&
A_{2}(\nu)
\left[\hat{\Phi}^\nu_c(r)\hat{\Phi}^\nu_c(r')\right]^{(n_1,n_2)},\cr
\G_{\ell m\omega}^{3,(n_1,n_2)}(r,r')&\equiv&
A_{3}(\nu)
\left[\hat{\Phi}^{-\nu-1}_c(r)\hat{\Phi}^{-\nu-1}_c(r')\right]^{(n_1,n_2)}.
\label{eq:G123}
\end{eqnarray}
\begin{figure}[b]
\begin{center}
\includegraphics[scale=.75]{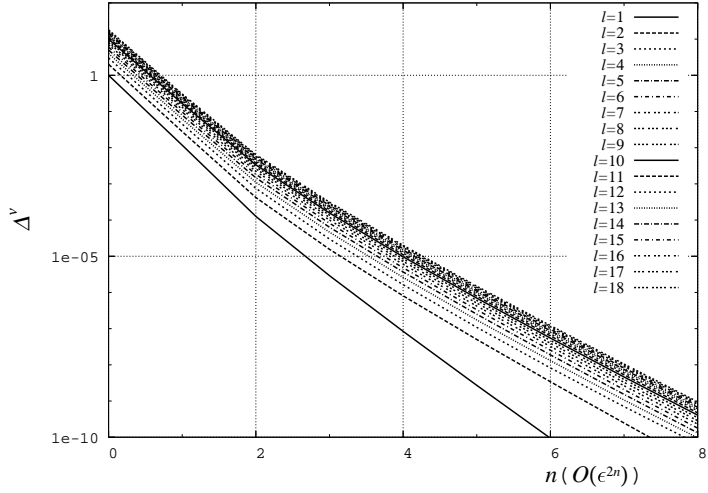}
\caption{ The error $\Delta^{\nu}(n)$ in $\nu$.}
\label{appfig:Coefferr1}
\end{center}
\end{figure}
\begin{figure}[b]
\begin{center}
\includegraphics[scale=.75]{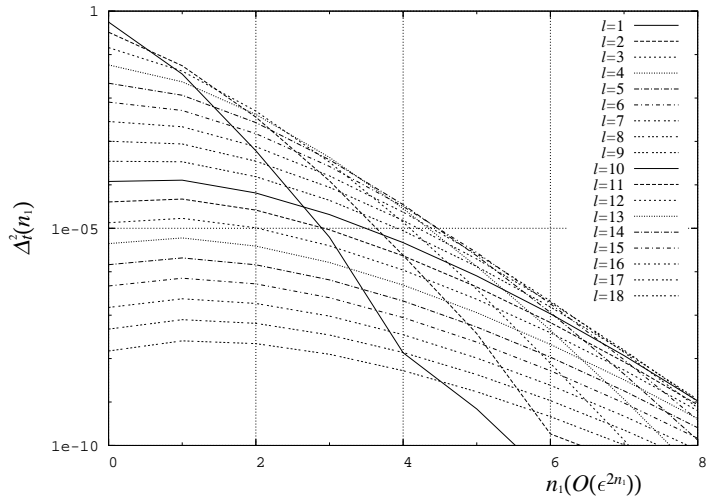}
\caption{Plots of the error in the
$t$-component of $F_\alpha^{2,(n_1,25)}$
for various values of $\ell$
as functions of the post-Minkowski order $n_1$.}
 \label{appfig:eps_err}
\end{center}
\end{figure}
\begin{figure}[b]
\begin{center}
\includegraphics[scale=.75]{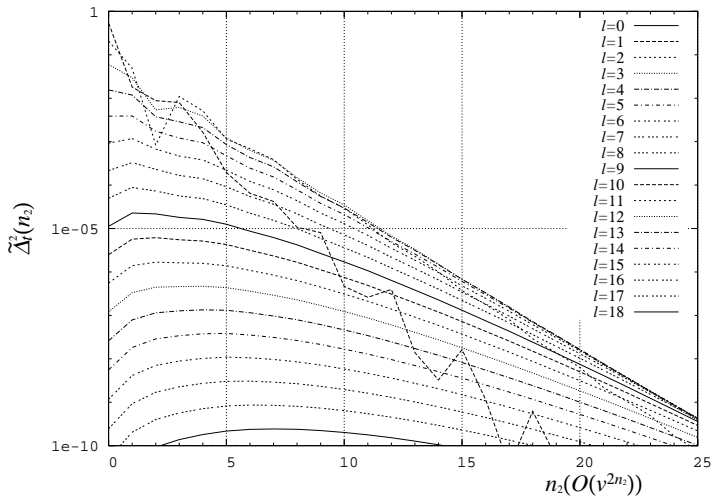}
\caption{Plots of the error in the $t$-components of
$F_\alpha^{2,(8,n_2)}$ for various values of $\ell$
as functions of the PN order $n_2$.} 
\label{appfig:pn_err}
\end{center}
\end{figure}
The indices $(n_1,n_2)$ indicate that the corresponding quantities are 
expanded in terms of $\epsilon$ and $v^2$ to $O(\epsilon^{2 n_1})$ and
$O(v^{2n_2})$, respectively. Note that $n_1$ and $n_2$ correspond to the
post-Minkowski order and post-Newtonian order, respectively.
The coefficients $A_{i}$ are composed of $\tilde{\beta}$,
$\tilde{\gamma}$, 
$W_{\ell m\omega}(\phi_{\rm c}^{\nu},\phi_{\rm c}^{-\nu-1})$,
$\epsilon^{\nu}$ and $U^{\nu}$. As noted above, because their accuracy
is essentially determined by the accuracy of $\nu$, we regard the 
$A_i$ as functions of only $\nu$, and we estimate the error
caused by using an approximate value of $\nu$.

First, we investigate the convergence of $\nu$,
which is given by a power series in $\epsilon$.
We define $\Delta^{\nu}(n)$ by
\begin{eqnarray}
\Delta^{\nu}(n)&\equiv& \left|[\nu]^{(n)}-[\nu]^{(n-1)}\right|.
\end{eqnarray}
Here, the index $(n)$ indicates that these quantities are expanded in
terms of $\epsilon$ to $O(\epsilon^{2n})$. Considering the most
relativistic case, we set $r=6M$ and $\epsilon=2M\ell\Omega$, where
$\Omega$ is the orbital angular velocity. We plot $\Delta^\nu(n)$ for
various values of $\ell$ in Fig.~\ref{appfig:Coefferr1}.
We see that the convergence is very fast.
Because the errors in $A_{i}$ are proportional to that in $\nu$,
we may ignore the errors in $A^{(i)}$ if we take, say, $n=8$.

We now investigate the convergence of the $\tilde R$-force.
Let us denote the part of the force that is due to  
$\G_{\ell m\omega}^{i,(n_1,n_2)}(r,r')$ by 
$F^{i,(n_1,n_2)}_\alpha$. As seen from Fig.~\ref{appfig:FtilRerr},
the $\tilde R$-force is dominated by $F^{2,(n_1,n_2)}_\alpha$.
This is in agreement with the order-counting carried out
previously~\cite{HJNSST}.
We define the errors $\Delta^{2}_\alpha(n_1)$ and 
$\tilde{\Delta^{2}}_\alpha(n_2)$ by
\begin{eqnarray}
 \Delta^{2}_\alpha (n_1)&\equiv&
\left|  \frac{F^{2,(n_1,n_2)}_\alpha - F^{2,(n_1-1,n_2)}_\alpha}
{F^{R}_\alpha}\right|,\cr
 \tilde{\Delta}^{2}_\alpha (n_2)&\equiv&
\left|  \frac{F^{2,(n_1,n_2)}_\alpha - F^{2,(n_1,n_2-1)}_\alpha}
{F^{R}_\alpha}\right|.
\end{eqnarray}
The $t$-components of $\Delta^{2}_\alpha(n_1)$ at $n_2=25$ and
$\tilde{\Delta^{2}}_\alpha(n_2)$ at $n_1=8$ are plotted in
Fig.~\ref{appfig:eps_err} and Fig.~\ref{appfig:pn_err}, respectively.
We see very fast convergence. Although not shown here, we find that 
the same is true for the other components.
These results confirm that our calculations are sufficiently accurate.

\end{document}